\documentclass[12pt,preprint]{aastex}
\begin{document} 
\shorttitle{SNU Bright Quasar Survey}
\shortauthors{Lee et al.}

\title{Seoul National University Bright Quasar Survey in Optical (SNUQSO) I: First Phase Observations and Results}

\author{Induk Lee\altaffilmark{1,6}, Myungshin Im\altaffilmark{1,7},
  Minjin Kim\altaffilmark{1}, Eugene Kang\altaffilmark{1},
  Hyunjin Shim\altaffilmark{1}, Gordon T. Richards\altaffilmark{2},
  Alastair C. Edge\altaffilmark{3},
  Myung Gyoon Lee\altaffilmark{1}, Changbom Park\altaffilmark{4}, and Myeong-Gu Park\altaffilmark{5}}

\altaffiltext{1}{Department of Physics and Astronomy, FPRD, Seoul National University, Seoul 151-747, Korea}
\altaffiltext{2}{Department of Physics, Drexel University, 3141 Chestnut Street Philadelphia, PA 19104}
\altaffiltext{3}{Institute for Computational Cosmology, Durham University, South Road, Durham DH1 3LE, UK}
\altaffiltext{4}{Korea Institute for Advanced Study, Dongdaemun-gu, Seoul 130-722, Korea}
\altaffiltext{5}{Kyungpook National University, Daegu 702-701, Korea}
\altaffiltext{6}{idlee@astro.snu.ac.kr}
\altaffiltext{7}{mim@astro.snu.ac.kr}

\newcommand{\oiii}{[$\ion{O}{3}$]}
\newcommand{\hb}{$H\beta$}
\newcommand{\ha}{$H\alpha$}
\newcommand{\newqso}{SNUQSO J003236.59-091026.2}
\begin{abstract}
   We present results from the first phase of
  the Seoul National University Bright Quasar Survey in
  Optical (SNUQSO) as well as its basic observational setup.
   Previous and current large area surveys have been successful in
  identifying many quasars, but they could have missed bright quasars 
  due to their survey design. 
   In order to help complete the census of bright quasars,
  we have performed spectroscopic observations of new bright quasar candidates
  selected from various
  methods based on optical colors, near-infrared colors, radio, and X-ray data.
   In 2005/2006, we observed 55 bright quasar candidates using
 the Bohyunsan Optical Echelle Spectrograph (BOES)
 on the 1.8 m telescope at the Bohyunsan Optical Astronomy
 Observatory in Korea.
   We identify 14 quasars/Seyferts from our observation,
 including an optically bright quasar with $i=14.98$ mag
 at $z=0.092$ (SDSS J003236.59-091026.2).
  Non-quasar/Seyfert objects are found to be mostly stars,
 among which there are 5 M-type stars and one cataclysmic
 variable. 
   Our result shows that there still exist bright quasars to be discovered.
 However, at the same time, we conclude that finding new bright quasars
 in high Galactic latitude regions is very challenging
 and that the existing compilation of bright quasars is nearly complete in
 the northern hemisphere.
\end{abstract}

\keywords{galaxies: active - galaxies: nuclei - techniques: spectroscopic - quasars: individual (SNUQSO J003236.59-091026.2) - quasars: emission lines - surveys}

\section{Introduction}

  Since the first discovery of a quasar more than 40 yr ago
 (Schmidt 1963), 
   many quasar surveys have been conducted to date, and the number of 
  known quasars and Seyferts is now well over 100,000
  (V\'eron-Cetty \& V\'eron 2006; Croom et al. 2004; Schneider et al. 2007). 
   Among numerous quasars discovered to date, bright quasars receive special 
  attention as an attractive astrophysical tool. Since they are bright, 
  one can investigate their intrinsic properties
 --- such as the black hole mass
  and accretion rate --- in detail,
  through long-term monitoring programs with small telescopes
  or getting high-resolution spectra
 (e.g., Kaspi et al. 2000; Hawkins 2007)
  By identifying more bright quasars, we can also possibly constrain
 the low-redshift luminosity function of quasars better, which can be 
 useful for understanding the cosmological evolution of luminous quasars at
 high redshift.

   There are many different techniques to find quasars, including 
 UV-excess (UVX) method (Palomar-Green survey; PG),  
 emission-line surveys using objective prisms 
 (Hamburg Quasar Survey; HQS hereafter; Hagen et al. 1995),  
 radio surveys (FIRST), and X-ray surveys.
 Here we describe a few major surveys related to bright quasars.

  Optical surveys target point-like sources with UV excess (UVX) as quasar 
 candidates. Quasars have a power-law spectrum, and the spectrum peaks at UV,
 which is called the UV bump. These unique spectral features distinguish quasars 
 from normal stars.  A representative large-area, bright 
 quasar survey is the Palomar Bright Quasar Survey (BQS). BQS targeted 
 1715 UVX stellar objects with $U-B < -0.44$ mag and $B < 16.6$ mag chosen
 from a $10,714~ {\rm deg^2}$ area of the northern sky covered by the Palomar Green
 Survey (PG; Schmidt \& Green 1983).
  Among these 1715 objects, 92 (5.36\%) were identified as quasars
 (Greene et al. 1986). 
  More recently, large area galaxy surveys such as the Sloan Digital Sky
 Survey (SDSS)  and the 2dF and 6dF surveys (Croom et al. 2004;
 Schneider et al. 2007) have discovered many quasars by selecting
 candidates mainly using multi-color selection criteria
 (Richards et al. 2002, 2004 ; Smith et al. 2005).
   The SDSS has improved optical quasar selection techniques
 by adopting sophisticated multi-color selection criteria (Richards et al.
 2002, 2004). Richards et al. (2004) show that their quasar selection
 technique efficiency is more than $\sim$65\%. Covering nearly the entire 
 northern hemisphere, SDSS has found more than 70,000 quasars so far
 (Adelman-McCarthy et al. 2005).

  Some other large-area bright quasar surveys--the Hamburg Quasar Survey
 (HQS, Hagen et al. 1995)
 and the Hamburg/ESO Survey (HES)--use objective prisms to identify 
 emission line objects. 
   HQS aims to cover $\approx 11,000 {\rm ~deg^2}$ of the northern sky
 with digitized objective prism plates. HES plans to cover
 a similar area ($9000 ~{\rm deg^2}$) of the southern hemisphere with 
 a similar technique. Both surveys target 
 bright objects ($13 < B <18$ mag for the northern hemisphere),
 and the HQS has found 343 bright quasars so far, 
 137 at $B\le 17 ~{\rm mag}$ (Hagen et al.1999).
  Fainter quasars are also discovered by 
 other slitless spectroscopic surveys (e.g., Large Bright Quasar Survey by
 Hewett et al. 2001).

  Bright quasars are also found by matching radio detections with  
 optical point sources.  As the name ``quasar'' stands for
 a radio source with stellar-like appearance, point sources
 with radio emission have a high probability of being a quasar.
 One of the largest radio-selected, bright
 quasar surveys is the FIRST Bright Quasar Survey (FBQS), which  
 selects radio sources with compact optical appearance as quasar
 candidates. FBQS reports the discovery of 957 such objects
 (White et al. 2000 ; Becker et al. 2001).

   These surveys cover significant portions of the sky, but
 it is quite possible that these surveys do not provide a complete
 census of bright quasars.
 Although SDSS has discovered nearly 100,000 quasars 
 with its efficient survey strategy (Schneider et al. 2007),
  SDSS cannot discover very bright quasars because of technical
 difficulties related to the survey design. When spectroscopically
 observing targets selected from photometric data, SDSS imposes a magnitude limit
 at $i=15.0$ mag, since the light from objects brighter
 than $i < 15$ mag spills over to neighboring fibers and ruins spectra
 of other targets.
  This leaves  BQS, HQS, and FBQS as major sources of very bright quasars,
 but these surveys may have their own problems.
  Many works have been carried out to see if there are any systematic biases 
 in BQS (Wampler \& Ponz 1985; Goldschmidt et al. 1999;
 Miller et al. 1993).
  Such studies find that the photometric inaccuracy in BQS 
 leads to omission of some bright quasars in their sample. For example,
 a recent work by Jester et al. (2005) shows that the photometric
 inaccuracy of the scanned photographic plate data used by BQS 
 moves the actual $U-B$ cut to $U - B < -0.71$
 from $U-B < -0.46$ as originally designed.
  Coupled with a photometric uncertainty of $\sigma_{U-B} = 0.24$ mag
 in the plate color, BQS seems to miss as many as 50\% of bright quasars
 satisfying the magnitude cut.
  Although shown to be fairly complete (Wisotzki et al. 2000), 
  an extensive search for bright QSOs might reveal QSOs missing
 from prism surveys.  
  Radio surveys seem to be an effective way  to find bright quasars, 
 as White et al. (2000) show that  most of bright quasars ($B < 16$)
 are also in the FIRST radio sources (Gregg et al. 1996;
 Becker et al. 2001). However, there might be bright quasars that are
 too radio-quiet to satisfy the FIRST detection.

  In order to build  more complete census of bright quasars ($i \lesssim
 15$ mag), we have begun a bright quasar survey in the optical, called 
 the Seoul National University Bright Quasar Survey in
 Optical (SNUQSO). Our survey plan is to use various quasar survey  
 techniques (optical multi-color, near-infrared photometry,
 radio/X-ray) to search for any missing bright quasars from existing
 extensive quasar surveys.
  Our main scientific goals are (1) to discover any peculiar bright quasars; 
 (2) to build complete census on bright quasars; and
 (3) to conduct follow-up observations of bright quasars to study  
 host galaxy and SMBH properties through long-term monitoring observations
 and high-resolution imaging.

  In this paper, we report results from the first phase of 
 this survey, which consists of the observations of
 bright quasar candidates at high Galactic latitude,
 selected from the public SDSS DR3 and DR4 releases,
 as well as NIR, radio, and X-ray data.
  Our results include the discovery of a very bright quasar and 13 other
 quasars/Seyferts. The second phase of the survey concentrates on
 quasars at low Galactic latitude, and it is 
 currently ongoing. 
We will first describe the candidate selection method
 (section 2), followed by the observation and the data reduction procedure
 (section 3). In section 4, we will show the efficiency of
 the candidate selection method used in this study and describe
 the properties of the newly discovered quasar.
  Finally in section 5, we will summarize our findings and provide future
 plans.
 Note that we use throughout the paper 
 the 2MASS photometric system for NIR data, 
 and the AB magnitude system for SDSS magnitudes
 (see e.g., Fukugita et al. 1996).
 The current best-estimate of 
 cosmological parameters of $H_{0}=71\,km\,sec^{-1},~\Omega_m=0.27,$ and
 $\Omega_{\Lambda}=0.73$ (Spergel et al. 2003) is adopted when calculating
 the rest-frame quantities.

\section{Sample Selection}

  This section describes various selection methods we used.
  We selected quasar candidates from three different data sets.
 The main data sets we used for our candidate selection are
 the SDSS Data Release 3 and 4 (hereafter DR3 and DR4, respectively)
 photometric data, $ROSAT$ Bright Source Catalog (1RXS-B),
 NRAO VLA Sky Survey (NVSS; Condon et al. 1998)
 and Two Micron All Sky Survey (2MASS) Point Source Catalog (PSC).
 Also, we selected additional candidates from the Quasars.Org (QORG) catalog
 by Flesch \& Hardcastle (2004) in order to test the QORG quasar selection
 method, which is also described later in this section.
  Table 1 summarizes targets observed so far.
 Note that the photometric data, presented in Table 1
 is not corrected for Galactic extinction.  
 Galactic extinction is small [$E(B-V) < 0.1$ mag] for
 objects in Table 1
 and does not affect our results nor selection of targets much.

\subsection{Optical Multiple Color Method}

  As mentioned in \S  1,
 special characteristics of quasar SEDs place
 the majority of quasars in regions of color-color space away
 from stellar locus
 (e.g., Richards et al. 2002). The simplest example of the quasar
 selection based on color is the BQS UVX selection criterion of 
 $U - B < -0.46$ mag.

  More sophisticated methods have been
 applied to the SDSS quasar selection (Richards et al. 2004;
 Schneider et al. 2007). 
  Richards et al. (2002) isolate stellar locus in
 $ugriz$ multi-color space and select stellar objects outside the
 stellar locus as quasar candidates.
  Figure 14 in Richards et al. (2002) shows the location of quasars
 and stars in color-color spaces and how quasars can be effectively
 selected from such diagrams.
  More recently, Richards et al. (2004) have
 improved the quasar candidate selection using the ``kernel
 density method''.  
  We used this kernel density method on the SDSS DR3 data set to select
 bright quasar candidates. Note that these quasar candidates were  
 picked as candidates in SDSS originally but not observed
 spectroscopically because they are too bright ($i < 15$ mag). With this
 method, we identified 44 bright quasar candidates, of which 
 22 were observable in 2005 January. We call these candidates 
 group A candidates.

 To find more bright quasars that might be missing in the kernel density method,
 we have gone back to a more straightforward color selection approach as well
 and selected additional candidates from the following color region:
{\setlength\arraycolsep{2pt}
\begin{eqnarray}
0.0<&u^{\ast}-g^{\ast}&<0.7, \nonumber\\
-1.0<&g^{\ast}-r^{\ast}&<4.0, \nonumber\\
-0.1<&r^{\ast}-i^{\ast}&<0.6, \nonumber\\
{\rm and}~~-0.3<&i^{\ast}-z^{\ast}&<0.5. \nonumber
\end{eqnarray}}

  The asterisk ($\ast$) denotes PSF magnitude, which is a magnitude measured by
 a PSF fitting method (Stoughton et al. 2002). The PSF magnitude should represent 
 the total flux of a point source well.
 The above color box represents the region where many quasars can be found
 in color-color plots of Figure 7 in Richards et al. (2002).
 However, the above color box includes objects such as 
 white dwarfs (WDs), M stars, or WD$+$M pairs.
 To improve the selection efficiency, we excluded the regions of WDs and WD$+$M pairs
 as defined in Richards et al. (2002).

First, the exclusion region of WDs is,
{\setlength\arraycolsep{2pt}
\begin{eqnarray}
-0.8<&u^{\ast}-g^{\ast}&<0.7, \nonumber\\
-0.8<&g^{\ast}-r^{\ast}&<-0.1, \nonumber\\
-0.6<&r^{\ast}-i^{\ast}&<-0.1, \nonumber\\
{\rm and}~~ -1.0<&i^{\ast}-z^{\ast}&<-0.1,\nonumber
\end{eqnarray}}
and the exclusion region of WD$+$M pairs is
{\setlength\arraycolsep{2pt}
\begin{eqnarray}
-0.3<&g^{\ast}-r^{\ast}&<1.25, \nonumber\\
0.6<&r^{\ast}-i^{\ast}&<2.0, \nonumber\\
{\rm and} ~~0.4<&i^{\ast}-z^{\ast}&<1.2. \nonumber
\end{eqnarray}}

 The SDSS photometry becomes inaccurate due to saturation for very bright 
 point sources ($i < 14$ mag).
 Also, as we go to the brighter magnitude, the ratio of bright stars versus 
 quasars satisfying the color selection criteria increases dramatically.
  As a consequence, the probability of a candidate being a
 quasar drops rapidly toward the brighter magnitude.
  Hence, we discarded candidates brighter than  $i=14$ mag.
  Through these processes, we selected about a 
 hundred additional candidates for observation in 2005 January.
 We name these additional candidates group B candidates.
 Figure 1 shows the color boxes of group B candidates, WD, WD$+$M pairs,
 and the stellar locus. 
 Note that the photometric data used in Figure 1 is not corrected for Galactic extinction as in Table 1.

\subsection{Optical and Near IR}

  A number of studies suggest that quasars tend to have red colors in NIR,
  e.g., $J-K\gtrsim2$ mag, while most stars have $J-K < 0.8$ mag 
  (Cutri et al. 2000; Barkhouse \& Hall 2001;
  Glikman et al. 2006;  Maddox \& Hewett 2006; Chiu et al. 2007).
  These studies imply that NIR data can be used to help identify
 quasars among point sources. 
  
  Therefore, we used NIR along with the optical selection 
 to boost the selection efficiency of quasars during the 2005 May observation run. 
  For this purpose, we expanded the faint limit of the magnitude cut to
 $i=16$ mag,
 although this overlaps with the SDSS quasar observation.
  The $15 < i < 16$ candidates
 were not observed by SDSS spectroscopically at the time of our observation
 and served as a sample
 to test our optical + NIR color selection method.
  The optical + NIR color selection of quasar candidates was done via
 the following steps.
  First, we chose candidates using the optical multi-color method described
 in the previous section. In this case, we extended the color region a little to find more quasars.
 Then, we matched  the selected objects with 2MASS point sources
 with $J-K > 0.8$ mag  within $1\arcsec$ matching radius.
 This process yielded roughly 50 candidates, of which we observed 17.
 When observing these candidates,
 we have given more weights to objects with large $J-K$ values.

 \subsection{Radio + NIR Selection}
  
  It is well known that a point source with radio emission is likely to be 
 a quasar (e.g.,  Becker et al. 2001; Hewett et al. 2001).
  By combining both the radio information and NIR color selection method,
 we can possibly discover quasars with minimal contamination from stars
 and other Galactic objects. To achieve this goal, 
 we matched NRAO VLA Sky Survey (NVSS; Condon et al. 1998) radio sources
 against 2MASS objects with $J-K \gtrsim 1.4$ mag to select
 quasar/AGN candidates.
 Our test of the efficiency of such a method on SDSS quasars 
 show a high quasar/Seyfert identification efficiency of $\gtrsim 85$\%.
 We observed five such quasar candidates at 
 high Galactic latitude. They were observed in 2006 June/December runs
 together with 88 
 low Galactic latitude quasar/AGN candidates as reported in Im et al. (2007).  
 The detailed description of the sample selection appears in Im et al.
 (2007) and I. Lee et al. (2008, in preparation).

\subsection{X-ray Sources}
  It has been known that many quasars are X-ray sources 
 (e.g., Stocke et al. 1991).
   This means that X-ray selection can help select quasars.
 To select quasar candidates, we used the $ROSAT$ All-Sky Bright 
 Source Catalog (1RXS-B), for which the positional accuracy is 
 known to be $\sim 13 \arcsec$ (68\% confidence; Voges et al. 1999). 
   We matched X-ray sources in 1RXS-B against 2MASS point sources
 with $J-K > 1$ mag and  $J < 16$ mag within the matching radius of
 $10\arcsec$ and chose them as quasar candidates.
   Therefore, this selection method is
  based on X-ray + NIR data.

\subsection{QORG catalog}

  As additional targets, we chose objects having a high 
 probability of being quasars in the QORG catalog of 
 Flesch \& Hardcastle (2004). Flesch \& Hardcastle 
 matched radio and/or X-ray sources to optical sources in 
 the APM/USNO-A catalogs. Then, they assigned a possible optical
 match to the X-ray/radio detections based on the probability that 
 the optical match is a quasar.
 The probability is based on the color,  
 the appearance (extended vs. point), 
 and the local density of optical sources around the X-ray/radio detection. 
 We observed six targets from the QORG catalog as filler targets.

\section{Observation and Data Reduction}

\subsection{Observation}
  We observed quasar candidates with the longslit, low resolution mode
 of Bohyunsan Optical Echell Spectrograph (LSS; Kim et al. 2003)
 on the 1.8 m telescope at the Bohyunsan Optical Astronomy
 Observatory (BOAO), Korea,
 over four observing runs spanning 
 from 2005 January to 2006 December.
  Table 2 summarizes the weather conditions and observing 
 parameters of each run.

  For our spectroscopic observation, we used a 150 groove mm$^{-1}$ grating
 and a slit with a width of 250 $\mu m$. The width and the length of
 the slit  correspond to
 $3.6\arcsec$ and $3.6\arcmin$ on the sky, respectively.
  This spectroscopic setup provides a wide wavelength coverage of 
 $\Delta \lambda \simeq 5000~ {\rm \AA}$, 
 and 5.2 \AA/pixel with a resolution
 of $\lambda/\Delta\lambda \sim 366$ or $\delta v \sim 
 820~ {\rm km~sec^{-1}}$ in full wIdth at half-maxium (FWHM). 
 The instrument spectral resolution has been measured from
 the comparison lamp spectra.
  In order to cover as much wavelength as possible for 
 identifying multiple lines for redshift identification, 
  we have chosen to use rather low-resolution but 
 wide-wavelength coverage as the spectroscopic setup.
  With our spectroscopic setup, we can detect \ha~ ($6563~\AA$) 
 from $z=0$ to $z=0.5$, \hb~ ($4861~\AA$)  
 from $z=0.028$ to $z=1.0$, $\oiii ~ 5007~\AA$  from 
 $z=0$ to $z=0.99$, and  [\ion{O}{2}] $3727~\AA$ doublet from $z=0.342$ to $z=1.683$.
  For high-redshift objects,  \ion{Mg}{2} ($2798 \AA,~ z= 0.787$ to $2.574$) and
  \ion{C}{4} ($1549 \AA,~ z=2.228$ to $5.456$) are useful redshift indicators.
  The slit width matches the poor seeing condition of 
 the BOAO site. 
  The spectrum is dispersed on a 1024-by-1024 CCD,
 whose spatial plate scale is 0.73\arcsec pixel$^{-1}$.
A detailed description of BOES-LSS system in BOAO is given in Kim et al. (2003).

  The weather during the observing runs was not very good,
 but we managed to observe 55 candidates in total.
  Typically, exposures were taken twice for each candidate, with 
 about 5-10 minutes for each exposure depending on the brightness
 of the target and the sky condition.  We shifted the position of the
 target on the slit by about $1\arcmin$ before taking the 
 second exposure for each candidate. When there was another point source
  near the candidate, if it
 had a similar color to the candidate, we tried to observe it also.
  The spectra of  the comparison lamp, FeNeArHe, were taken right before and after
 each exposure for the wavelength calibration.
  We also observed several spectrophotometric standard stars
 for flux calibration.
  We used tungsten halogen lamps (THL) for a flat field.

\subsection{Reduction}

   We reduced all the data using IRAF\footnote{``IRAF is
  distributed by the National Optical Astronomy Observatory, which
  is operated by the Association of Universities for Research in
  Astronomy, Inc., under cooperative agreement with the National
  Science Foundation."} packages.
   The data were reduced with the standard procedures of bias subtraction,
  flat-fielding, aperture extraction, wavelength calibration,
  and flux calibration.

  Since the seeing condition was not stable each night and some of the
 nights were
  non photometric, the above flux calibration includes some inaccuracies.
  To improve the flux calibration, we renormalized 
  the spectrum for a newly discovered quasar (see next section). 
  This renormalization is done by adjusting the flux-calibrated spectrum
 to match the known photometric information of the object.
  For this purpose, we used the SDSS photometric data.
  Comparison of the SDSS photometry and the spectrum 
  with the original flux calibration reveals that the flux of the spectrum 
 is underestimated compared to the SDSS photometry by 8.1\%, 19.8\%,  and 25.0\%
 in the $g, r,$ and $i$ bands, respectively. 
  To reduce the difference, we recalibrated the flux of the spectrum to match
 the SDSS photometry at $r$ band. 
  This comparison shows that the flux calibration is accurate for
 the renormalized quasar spectrum at about the 10\% level.
  We have not performed renormalization of the flux calibration for 
 the other objects, but the above exercise suggests that the flux value is 
 accurate to about 25\%.

  Finally, we measured the redshift by running the `xcsao' task of `rvsao'
 package.
 The observed wavelengths of the \ha, \hb, and [OIII] are matched with the 
 emission line templates, and the recession velocity of each object
 is determined by a cross-correlation technique.

\section{RESULTS}

   We observed 17, 17, 5, 10, and 6 quasar candidates from the optical
  multi-color selected  sample (section 2.1), the optical+NIR
  color selected sample (section 2.2), the radio+NIR color selected sample 
 (section 2.3), the X-ray+NIR color selected sample (section 2.4),
  and the QORG sample (section 2.5),
  respectively.
   Among these objects, we confirmed that 1, 3, 4, 6, and 0 are quasars.
  Spectroscopic properties of these 14 quasars/Seyferts are presented
 in Table 3.
  Some of them were also discovered by SDSS after our observation, and
 such cases are marked in Table 1 and 3.

\subsection{Newly Discovered Quasars}

  Fig. 2 and Fig. 3 show the flux calibrated spectra and images 
 of newly discovered quasars/Seyferts.
  The spectra in Fig. 2 show strong [OIII] emission lines or broad emission
 lines which are characteristic of quasars/Seyferts.
  Table 3 lists the parameters of each quasar such as redshifts,
 coordinates, FWHMs, and absolute magnitudes for the $g$ and $i$ bands.
  We classify objects as quasars when their Balmer line FWHM widths 
 exceed 1000 $km\,s^{-1}$ and their absolute magnitudes are 
 brighter than $M_{i} = -22$ AB mag (Schneider et al. 2003).
 Three objects, J143008.65+230621.6, J153334.69+051311.4, and J160731.42+173138.4 in Table 3, are found to
 have $M_{i} > -22$ AB mag, and therefore we classify these objects as Seyferts.
  One exception is the object J112632.90+120438.0, for which an emission line was marginally detected.
  This object was later identified as a broad-line quasar at $z=0.977$ by SDSS. 
  Therefore, we include this object in the list of quasars.

   The line widths are measured for two emission lines, 
 H$\alpha$ and H$\beta$, using a single-component Gaussian-fitting,
 instead of a two-component profile of 
 narrow$+$broad lines (e.g., Kim et al. 2003),
  since the spectral resolution is not high enough to
 effectively  separate the two components.
  Note that this procedure can underestimate the broad line width.
 The continuum of the spectrum under each emission line was determined
 by a straight line interpolation between two points that lie sufficiently
 far from the emission line.
  We tried several different sets of two points to see how the FWHM values
 change,  but did not find any significant variation for \ha.
  The above procedure did not work as well for \hb~ since the \hb~ lies
 in a part of the spectrum where calibration is in suspect
 (see section 4.2). The instrument resolution was also corrected for the
 line width measurements.

  When deriving the absolute magnitudes, we used a $K$-correction
 assuming a power law of $f_\nu \propto \nu^\alpha $, ignoring host galaxy and
 emission line contribution (see \S $5.4.$).
  In such a case, $K$-correction can be expressed as a simple form of
 $K(z)=-2.5(\alpha+1)\log(1+z)$, and we used the value of $\alpha=-0.5$ (Schmidt \& Green 1983; Boyle et al. 1988).
  Note that for objects where the host galaxy is clearly visible (J102700.00+390804.2, J103858.44+401058.4, J124127.66+085219.5, 
 J143008.65+230621.6, J145434.35+080336.7, J153334.69+051311.4, and J160731.42+173138.4),  
 the absolute magnitudes in Table 3 serve only as a rough measure,
 since we ignored the host galaxy light in calculating the absolute magnitudes.
  Also we estimate that the contribution
 of \ha ~ to the total flux in $i$ band can be as much as $\sim$25\%. 

  The most notable discovery is \newqso, which 
 ranks among the brightest quasars discovered to date
 [$i=14.98$ mag or V=15.3mag, adopting the relation of $V=g-0.52(g-r)-0.03$  
 in Jester et al. 2005].
  To determine where this quasar ranks in terms of apparent brightness,
  we matched \newqso~ against a list of
 known quasars searched in the NASA/IPAC Extragalactic Database (NED) 
 \footnote{http://nedwww.ipac.caltech.edu} .
  A simple NED search reveals that this quasar corresponds to
 the 5th--14th brightest at $V$-band among bona fide ``pointlike''
 quasars (as of 2007 July 16).
  In any case, the discovery of \newqso~ suggests that there still exist 
 very bright quasars waiting to be discovered, although the number of
 such objects might be very small.

\subsection{Non-Quasars and Unidentified Objects}

   Fig. 4 shows the representative spectra of objects classified as
 stars, as well as interesting stellar objects such as M stars and
 a cataclysmic variable.
  We identified objects as stars if their spectra show H$\alpha$
 absorption line at  low radial velocity.
  Only one of the candidates (J105756.30+092314.9) other than quasar/AGNs showed a
 clear sign of an emission line at H$\alpha$. The velocity dispersion
 of this object is found to be $\sigma_v \simeq 375 km\,s^{-1}$, 
 and we classify this as a cataclysmic variable star from the spectral 
 feature and the identification of H$\beta$ and H$\gamma$ lines using
 additional data we obtained in 2005 February.
 In addition, we identify five M stars 
 based on their spectral shapes.
 However, except for J043737.4-022928A, they 
 are located very close to another brighter star,
 and the spectra of these stars might be
 contaminated by nearby stars.
 For other stars, we could not classify spectral types,
 because of the limited spectral coverage 
 we used at blue wavelengths 
 ($5000-10000{\rm \AA}$ or $4200-8400{\rm \AA}$)
 and concerns about the calibration at $5000-6000 {\rm \AA}$ where 
 an artificial bump appears possibly due to instrument defects.

\section{DISCUSSIONS}

\subsection{Quasar Selection Efficiency}

 Here we examine the quasar selection efficiency of various methods,
 which is summarized in Table 4.

\subsubsection{Optical Multiple Color Method}

  For the optical multiple color method, 
 we found only one quasar out of group A and B targets 
 (17 total).
  Therefore, the efficiency of our candidate selection method seems very low,
  which is consistent with Richards et al. (2006), who find 
 the efficiency of the optical multiple color
 ($ugri$) selection method at bright end ($i<15.5$ mag) is less than $5\%$. 
  To find ways to improve the bright quasar selection efficiency,
 we examine the property of targets in more detail.

   The newly discovered quasar, \newqso, is detected in both radio and IR.
   Two objects selected with this method are found to be 
 radio sources; therefore, the efficiency is 50\%
 if the selection was done using radio detection.
   Although this number is based on small number statistics,
 it is interesting to note that the efficiency is consistent with
 that of FBQS (e.g. White et al. 2000).

  In the case of nonradio, optically selected bright quasar candidates,
 none turned out to be quasars.
  Therefore, we conclude that it is very unlikely to find an 
 optically selected bright quasar that is not a radio source.
  We can think of two reasons for this low efficiency.
  The first reason could be stellar contamination.
   The surface density of quasars decreases toward bright magnitudes,
 while the number of stars does not change much.
   Therefore, the optical selection tends to pick up more stars than quasars 
 at bright magnitudes.
   Second, the majority of optically bright quasars are detectable
 in radio at the current radio survey limits.
  For example, Fig. 13 of White et al. (2000) shows that
 the surface density of radio-selected quasars is almost the same as
 that of optically selected quasars at bright magnitude,
 suggesting that the optical bright quasars have almost complete overlap with
 radio-selected quasars.

\subsubsection{Other Methods}

  For the optical+NIR color selection method, we find that the efficiency
 is 17.6\%.
   Three quasars identified with this method are found to have $J-K>1.5$ mag
 and $15<i<16$. This re-confirms the idea that point 
 sources with non-stellar optical colors and red, NIR colors
 are likely to be quasars.

  For the radio+NIR color selection method, 
  we find that the identifiication efficiency is quite high (80\%). This 
  is consistent with the value we obtain through simulations using 
 SDSS quasars (I. Lee, et al. 2008, in preparation). This is not surprising, 
 considering that this selection method is a combination of two efficient 
 quasar identification methods

   For the X-ray selected targets,
  we find the identification efficiency
 at its face value is about 60\%.
  This confirms that X-ray data are quite valuable for finding
 AGNs, especially when the data are combined with NIR colors.

   For QORG targets, we could not find any quasar or AGN.
   The quasar \newqso  could have been in the QORG catalog,
 but it appears that the quasar was mismatched to
 a star with $i=13.8$ mag, 6\arcsec~ away from \newqso.

\subsection{Radio Properties of New Quasars}

   Fig. 5 shows broadband SEDs of several new quasars from UV to radio 
 wavelengths. In Fig. 5, we also plot SEDs of well-known 
 radio-loud quasars. 
   The figure indicates that the newly discovered quasars have lower fluxes 
 at radio than typical radio-loud quasars.
  In order to measure the radio loudness, we calculated the radio loudness
 parameter $R^{*}$ for each QSO.
   The $R^{*}$ parameter is taken as the ratio of
  flux in radio at 6 cm versus optical flux at $4400~{\rm \AA}$, and
  for radio-loud quasars, the $R^{*}$ value exceeds 10 (Kellermann et al. 1989;
  Stocke et al. 1992).
  When calculating $R^{*}$ parameter, we adopt $g$-band flux as 
 the flux at 4400 \AA~ and the 6 cm flux is computed from the 21 cm flux assuming
 a power law of $f_{\nu} \sim \nu^{-0.46}$ (Ho \& Ulvestad 2001).
  When quasars do not appear in the NVSS catalog or in the FIRST catalog, 
 we derived upper limits of $R^{*}$ using the detection limit of
 the NVSS survey ($\approx 2.5$ mJy).

  The radio loudness of quasars is summarized in Table 3.
 We find that only one of them belongs marginally to the radio-loud category.
 Thus, we can conclude that the majority of the newly discovered QSOs are
 radio-quiet.

\subsection{Mass of Black Holes}

   We estimate the mass of supermassive black hole residing in quasars using
  the method of Greene \& Ho (2005) which uses $L_{{\rm H_{\alpha}}}$
 and $\sigma_{v}$ of $H_{\alpha}$ with, 

\begin{equation}
M_{BH}=2.0^{+0.4}_{-0.3} 
\times 10^6 \Bigg(\frac{L_{H\alpha}}
{10^{42}~{\rm erg~s^{-1}}}\Bigg)^{0.55\pm0.02}
\Bigg(\frac{{\rm FWHM}_{H\alpha}}
{10^3~ {\rm km~s^{-1}}}\Bigg)^{2.06\pm0.06} M_{\sun}.
\end{equation}

  The estimated $M_{{\rm BH}}$-values are presented in Table 3. 
  We find that the SMBH masses range between $10^{7}$ and $10^{9} M_{\sun}$.
  For \newqso, it is about $2.6^{+0.9}_{-0.9}\times 10^{7} M_{\sun}$ .

\subsection{Host Galaxies}

 In this section, we examine the properties of the host galaxies and
 their detectibility.

  We use the $\sigma_{v}$ and  
 $M_{\rm {BH}}$ correlation of Tremaine et al. (2002) to get  
 $\sigma_{v}$ and then obtain the host spheroid luminosity
 $L$ using the Faber-Jackson relation.

  Table 5 lists the host galaxy $\sigma_{v}$ and the luminosity.
  The host spheroid systems expect to have various mass properties
 in the range $\sigma_{v}$ = $60$--350 $km\,s^{-1}$. 
  
  In order to check the detectibility of the host galaxies,
 we examined their expected apparent brightnesses and sizes.
  The apparent magnitudes are estimated by adopting the
 
  $K$-correction of $K_{corr,gal}(z) = z$  mag for $i$ band,
 and the rest-frame color of $B - i = 1.75$ mag and $B-g = 0.5$
 mag of early-type galaxies (Fukugita et al. 1995).
  In order to estimate the apparent size, we calculate the $r$ band
 half-light radius,  $\bar{r}_{50}$
 from the $\bar{r}_{50} - M_v$ relation of McIntosh et al. (2005).

  The derived properties of host galaxies are listed in Table 5.
  It is interesting to compare the predicted host galaxy luminosity versus
 the observed host galaxy brightness for quasars for which 
 host galaxies are visible (J102700.00+390804.2, J103858.44+401058.4, 
 J124127.66+085219.5, J143008.65+230621.6,
 J145434.35+080336.7, J153334.69+051311.4, J160731.42+173138.4 in Fig. 3).
   Assuming that the fiber magnitude (3\arcsec diameter) represents
 the light from a bright, central
 region, we attempted to measure 
  a rough value of 
  the host galaxy luminosity by subtracting
  fiber magnitude from the Petrosian magnitude.
   The measured values mostly are brighter by up to a few magnitudes,
 or similar to the predicted host galaxy luminosities within a
 few tenths of a magnitude, which is not very surprising
 considering that the predicted host galaxy luminosities
 take into account the luminosity of spheroidal components only.
  One exception is J153334.69+051311.4, whose predicted magnitude
 is about 1 mag fainter than the Petrosian-Fiber magnitude.
  The Petrosian magnitude of this object is similar 
 to the predicted $i$ magnitude, therefore we suspect that the Fiber magnitude
 is not a good representation of the central point source in this case.

   For the other cases, the host spheroids seem to be not only much fainter
 (by a few magnitudes) than the quasars but also small in their apparent sizes.
   Therefore, finding an underlying host galaxy would require a high-resolution
 imaging with high contrast, and the apparent absence of the host galaxy image
 in the SDSS imaging data seems natural.

\section{Conclusion}

  We have carried out a bright quasar survey to discover new bright quasars 
 that have been missed in previous surveys. 
  Bright quasar candidates are selected from SDSS and 2MASS data in 
 several ways, including the optical multi-color method, 
 the optical+NIR color selection method, the radio+NIR color selection method,
 and the X-ray+NIR color selection method. Some candidates are also taken from
 the QORG catalog.
  Our spectroscopic observations of 55 targets at the BOAO observatory 
 reveal 14 new bright quasars, while most of the other targets are found
 to be stars including 1 cataclysmic variable and 5 M-types. 
  Among 24 optically selected
 or optical+NIR color selected $i < 15$ mag quasar candidates,
 only \newqso~ has been identified as a new quasar. 
  There still exist very bright
 quasars missing in previous surveys, although the number of such ``missed'' 
 bright quasars is probably very small.
  The use of multiwavelength data sets proves to be a more efficient way
 to find bright quasars. When applied to $i \lesssim 17$mag candidates,
 these selection methods uncovered 13 quasars out of 25  
 candidates. Six candidates from the QORG catalog are found to be 
 stars.  

   We have studied the properties of the new quasars as well.
  For \newqso, we estimate an SMBH mass of
  $2.6 \times 10^{7}~ M_{\odot}$, suggesting that the host galaxy of
 this quasar is similar to a dwarf elliptical M32, 
 the bulge of the Milky Way, or M32. The close proximity of this object to
 a neighbor star ($i=13$ mag, 6\arcsec ~apart) makes \newqso~ a good target 
 for adaptive optics observation from the ground.
  The SMBH mass of other quasars ranges from $10^{6}$ to  $10^{9}~M_{\sun}$.

   Despite our discovery, we find that the probability of finding
 very bright quasars ($i < 15$mag) not in existing catalogs is very low,
 and the previous list seems almost complete
 for bright quasars at high Galactic latitude regions
 in the northern hemisphere.

\acknowledgments

  This work was supported by grant R01-2005-000-10610-0 from the Basic 
 Research Program of the Korea Science and Engineering Foundation.
 Also, I.L., M.K., E.K., and H.S. are supported in part by
 the Brain Korea 21 program at Seoul National University.
 C.B.P and M.G.P. acknowledge the support of the Korea Science and
Engineering Foundation (KOSEF) through the Astrophysical
Research Center for the Structure and Evolution of the Cosmos
(ARCSEC). Funding for the SDSS and SDSS-II has been provided by the Alfred P. Sloan Foundation,
 the Participating Institutions, the National Science Foundation, the US Department of Energy, the National Aeronautics and Space Administration, the Japanese Monbukagakusho, the Max Planck Society, and the Higher Education Funding Council for England. The SDSS Web site is http://www.sdss.org/.
The SDSS is managed by the Astrophysical Research Consortium for the Participating Institutions. The Participating Institutions are the American Museum of Natural History, Astrophysical Institute Potsdam, University of Basel, University of Cambridge, Case Western Reserve University, University of Chicago, Drexel University, Fermilab, the Institute for Advanced Study, the Japan Participation Group, Johns Hopkins University, the Joint Institute for Nuclear Astrophysics, the Kavli Institute for Particle Astrophysics and Cosmology, the Korean Scientist Group, the Chinese Academy of Sciences (LAMOST), Los Alamos National Laboratory, the Max-Planck-Institute for Astronomy (MPIA), the Max-Planck-Institute for Astrophysics (MPA), New Mexico State University, Ohio State University, University of Pittsburgh, University of Portsmouth, Princeton University, the United States Naval Observatory, and the University of Washington. 
 We thank Sanggak Lee, Marcel Agueros, Don Hoard,
 and others for their useful inputs on stellar classification, 
 and Yoojae Kim and Kyungsook Chung for obtaining the spectrum of 
 the cataclysmic variable star discussed in the paper. 
 Also, we thank the staffs at BOAO, especially Kang-Min Kim and 
 Byeong-Cheol Lee for their professional aid during our observing run.

Facilities: BOAO 1.8m telescope, BOES LSS

\clearpage

\begin{figure}
\label{fig1}
\figurenum{1}
\plotone{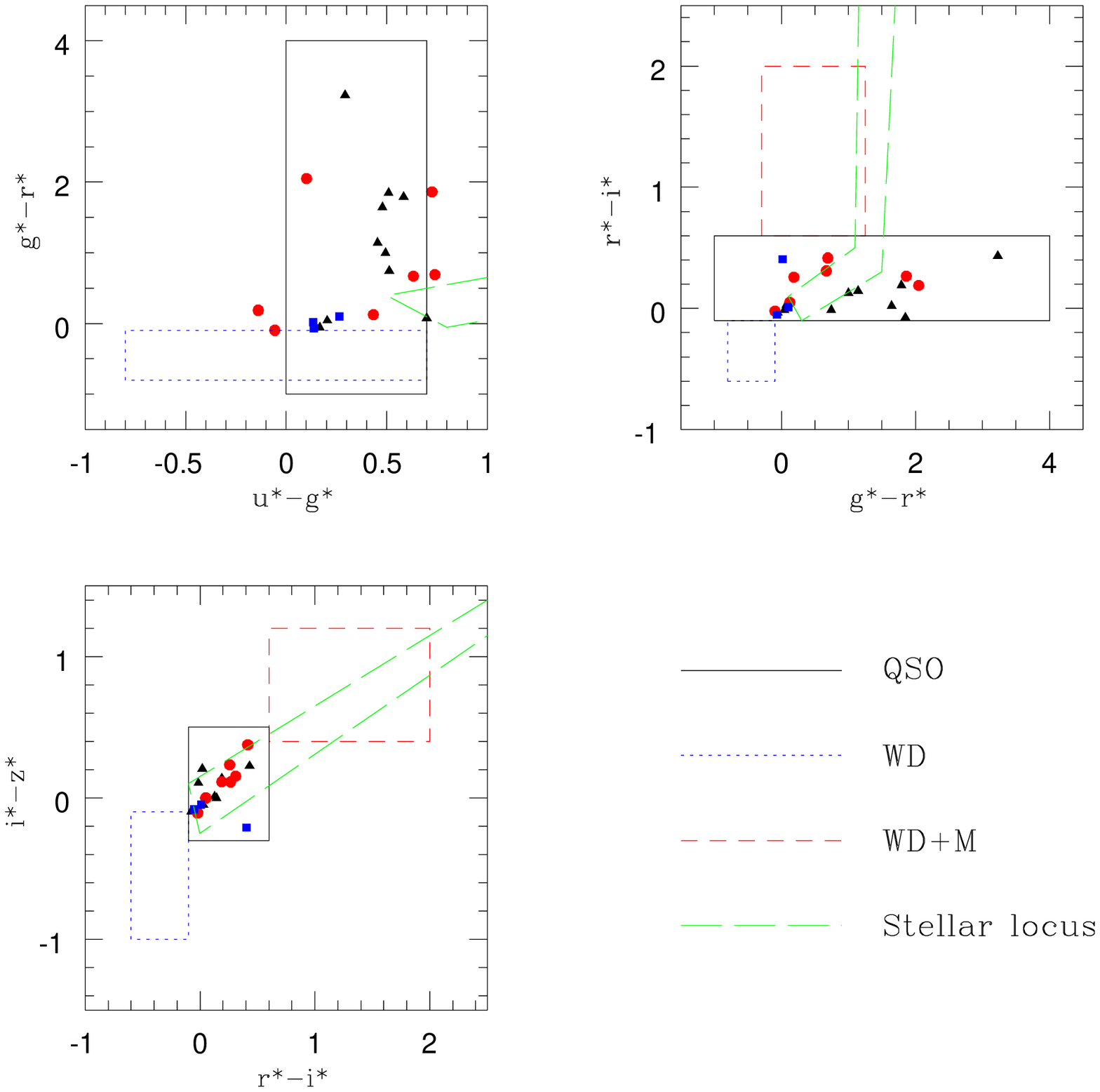}
\caption{Color-color plot of observed targets. Red circles and blue triangles indicate group A and group B candidates,
 respectively,  and squares show candidates that satisfy both group A and group B criteria.
 Black solid, blue dotted, and red short dashed boxes represent quasar, white dwarf, and white dwarf + M star regions,
 respectively. Green long dashed boundary shows location of the stellar locus.}
\end{figure}

\begin{figure}
\label{fig2}
\figurenum{2}
\plotone{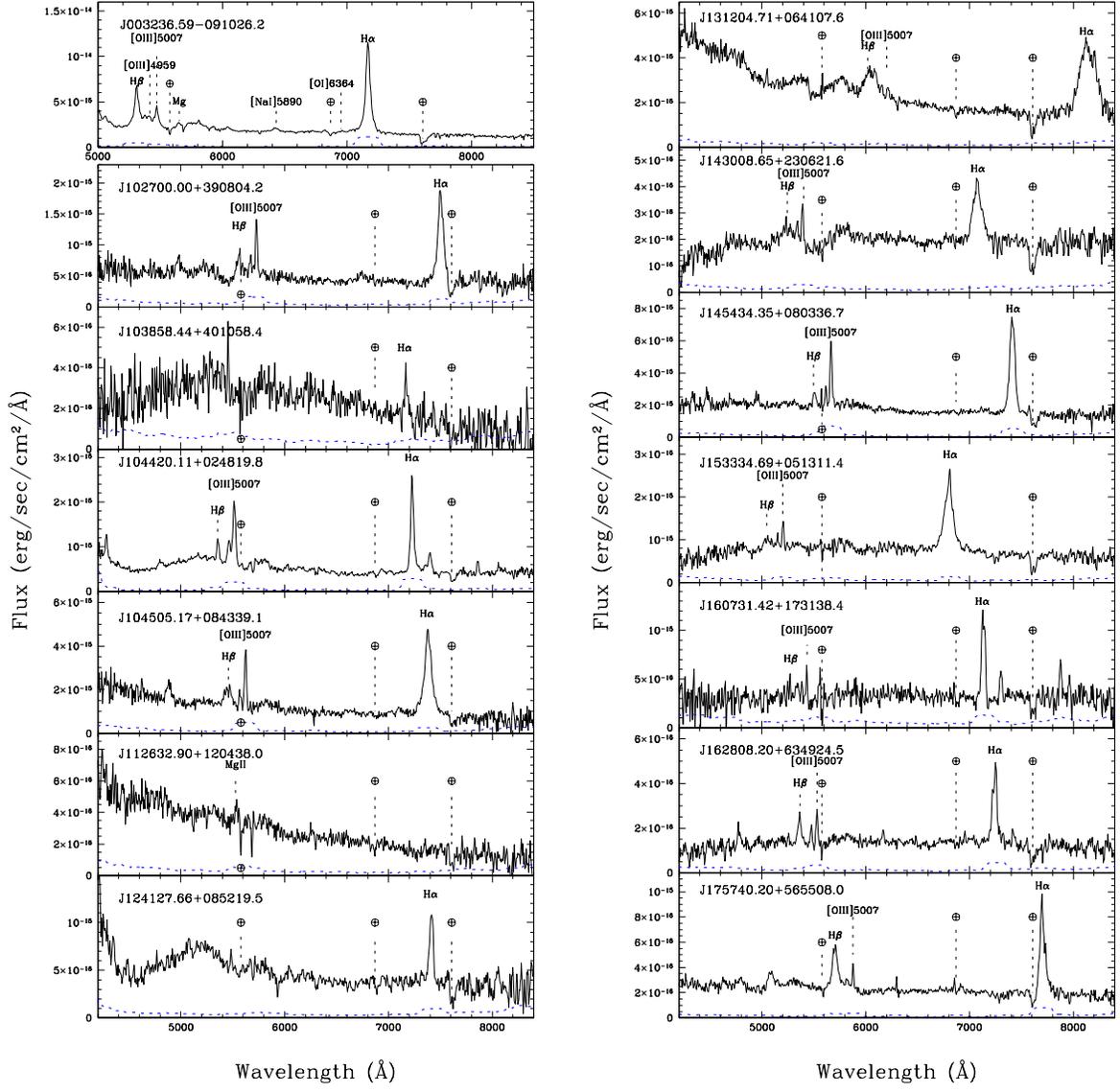}
\caption{Spectra of newly discovered quasars. The dashed lines indicate $1~\sigma$ error. Symbol $\oplus$ shows the positions of the sky absorption lines.}
\end{figure}

\begin{figure}
\label{fig3}
\figurenum{3}
\plotone{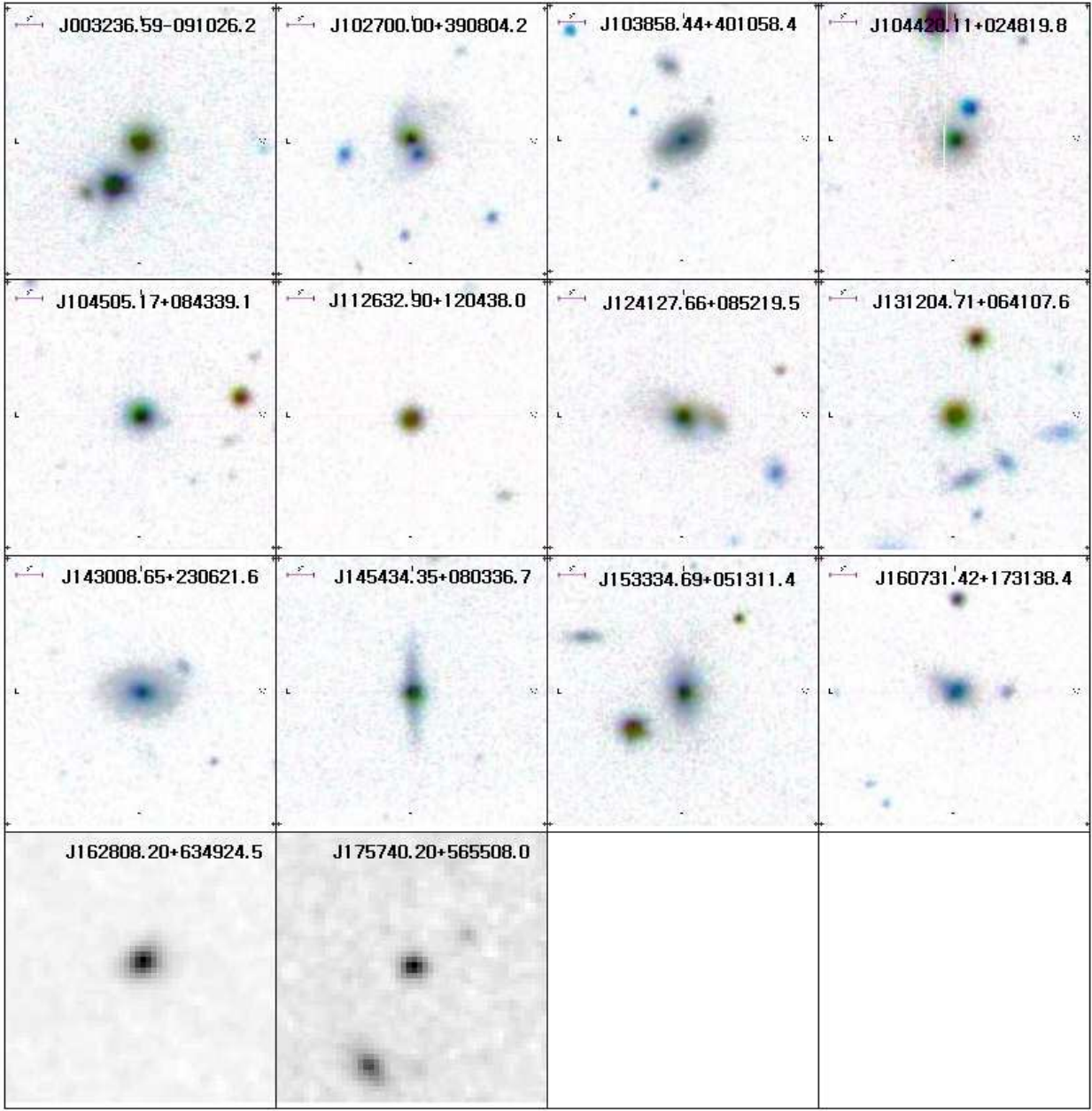}
\caption{$50\arcsec \times 50\arcsec$ images of newly discovered quasars.
 The images of 'J003236.59-091026.2' through `J153334.69+051311.4' are color-composites 
 of $g$ (blue), $r$ (green), and $i$ (red) images taken from the SDSS database.
 The images of `J160731.42+173138.4' through `J175740.20+565508.0' are from POSS-II red plate data.}
\end{figure}

\begin{figure}
\label{fig4}
\figurenum{4}
\plotone{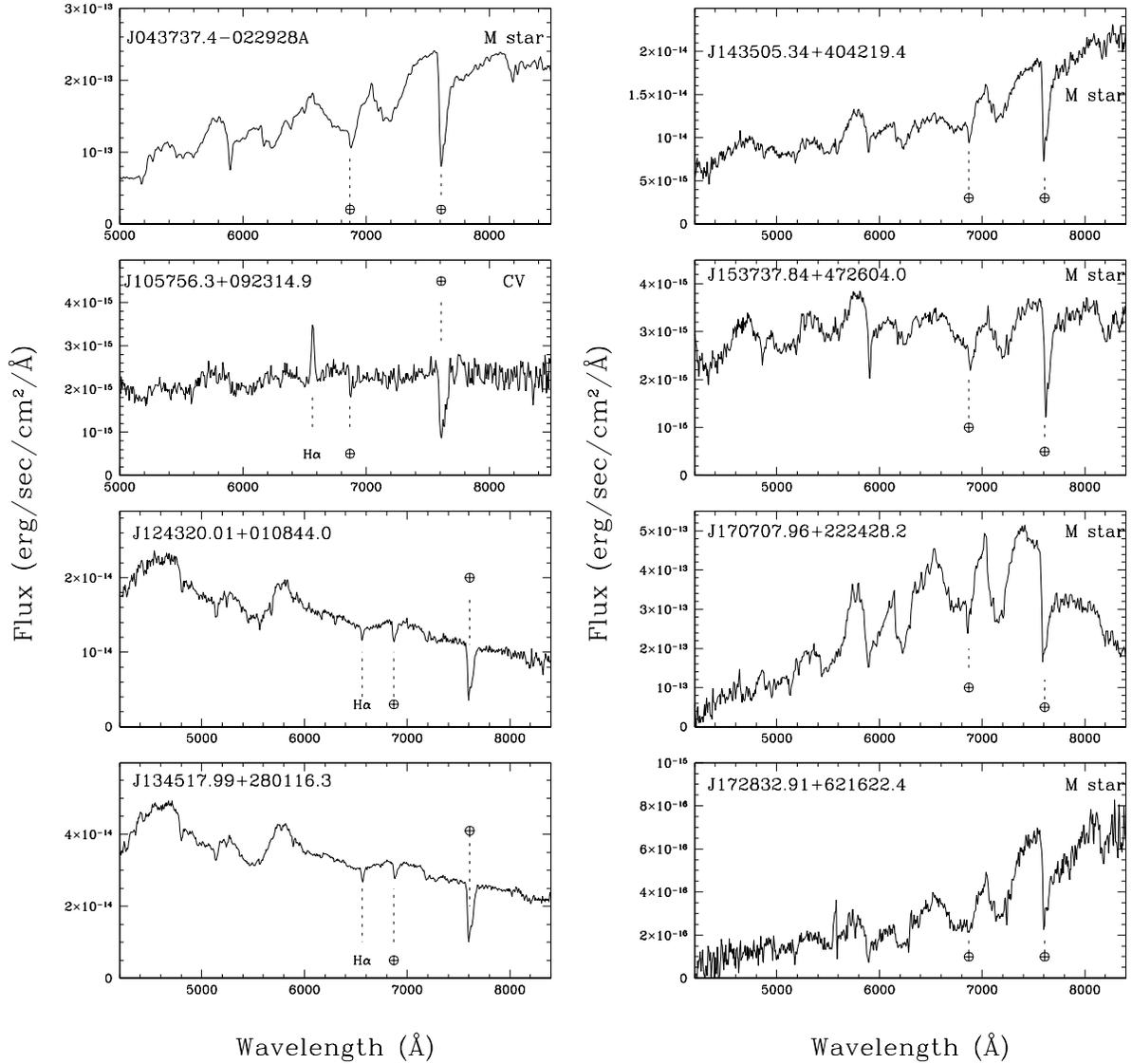}
\caption{Representative spectra of several objects identified as stars,
 including a cataclysmic variable and M stars.}
\end{figure}

\begin{figure}
\label{fig5}
\figurenum{5}
\plotone{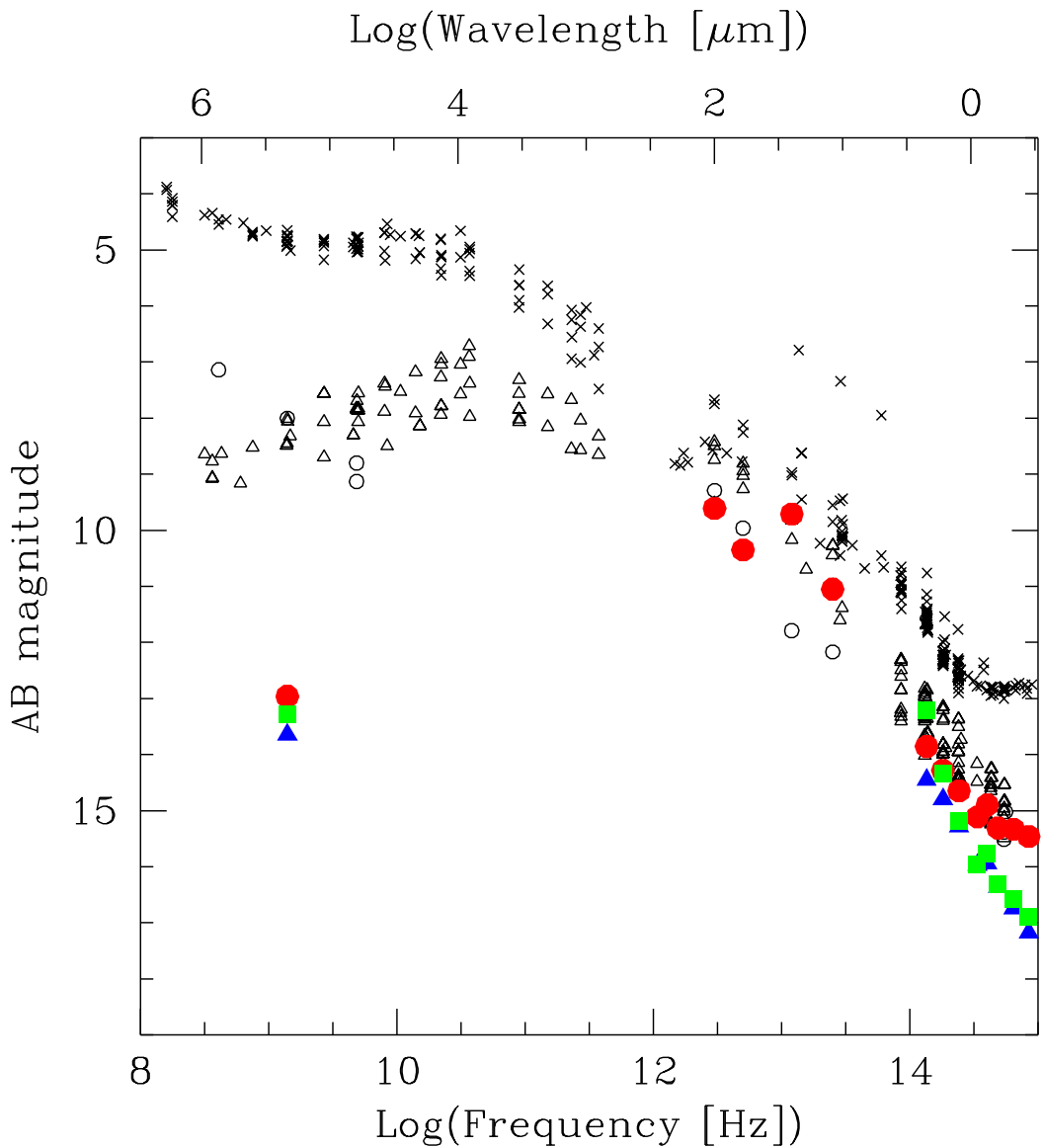}
\caption{Spectral energy distributions of 6 quasars.
 Crosses are for 3C273, open triangles for OJ+287,
 open circles for 3C066A,
 red filled circles for SNUQSO J003236.59-091026.2,
 blue filled triangles for SNUQSO J145434.35+080336.7,
 and green filled squares are for SNUQSO J104505.17+084339.1.
 The data of 3 quasars, 3C 273, OJ+287,
 and 3C 66A, are taken from NED.}
\end{figure}

\clearpage

\begin{deluxetable}{lllllcc}
\tablecolumns{7}
\tabletypesize{\scriptsize}
\tablewidth{5.5in}
\tablenum{1}
\tablecaption{Target summary}
\tablehead{
  \colhead{Name} &
  \colhead{Type\tablenotemark{a}} &
  \colhead{Method} &
  \colhead{$i$\tablenotemark{b}}&
  \colhead{$J-K$}&
  \colhead{Group}&
  \colhead{SDSS?\tablenotemark{c}}\\
}
\startdata
J000315.82+004017.8 &Star&	Optical 					&14.43	&0.41	&B   &No \\
J001002.11+004809.6 &Star&	Optical 					&14.46	&0.44	&B   &No \\
J003236.59$-$091026.2& Quasar&	 Optical 	&14.98			  	&1.99	&Both&Yes\\
J005229.4+001456   &Star&	QORG 					&     		&     	  &    &No \\
J012055.4+000454   &Star&	QORG 					&     		&     	  &    &No \\
J021505.57$-$091606.5 &Star&	Optical 					&14.09  &0.24	&B   &Yes\\
J025058.8+001226   &Unknown&QORG 					&           &     	 &    &No \\
J043737.4$-$022928A\tablenotemark{d}  & M Star &QORG 	&   		&     	 &    &No \\
J043737.4$-$022928B\tablenotemark{d}   &Star&	 QORG &	 			&     	&    &No \\
J052716.4+014840   &Star&	QORG 					&           &     	&       &No \\
J073845.46+171249.1 &Star&	Optical 					&14.94	&0.44 	&B   &No \\
J074532.64+413633.8 &Star&	Optical 					&14.59	&0.36	&B   &No \\
J075247.90+360110.4 &Star&	Optical 					&14.61	&0.29	&Both&Yes\\
J080520.22+064609.7 &Star&	Optical 					&14.92	&0.26	&B   &No \\
J082507.20+200637.6 &Star&	Optical 					&14.99	&0.23	&B   &Yes\\
J083121.17+475053.2 &Star&	Optical 					&14.06	&0.32	&B   &Yes\\
J085825.29+073944.9 &Star&	Optical 					&14.48	&0.52	&A   &Yes\\
J094141.91+572957.4 &Star&	Optical 					&14.52	&0.47	&A   &Yes\\
J094949.12+031616.6&Star&	 Opt.+NIR&13.95&2.32&&No\\              
J095810.81+014218.7 &Star&	Optical 					&14.61	&0.33 	&A   &No \\
J100505.03+431735.9 &Star&	Optical 					&14.59	&0.66	&B   &Yes\\
J102700.00+390804.2&Quasar&	 X-ray source&16.55\tablenotemark{e}&1.65&&Yes\\          
J103858.44+401058.4&Quasar&	X-ray source&15.91\tablenotemark{e}&1.15&&Yes\\
J104420.11+024819.8&Quasar&radio+NIR&16.28&2.00&&No\\
J104505.17+084339.1&Quasar&	 Opt.+NIR&15.76&1.98								&&Yes\\              
J105756.30+092314.9 &  CV\tablenotemark{f}&Optical &14.94		&0.23	&A   &Yes\\
J110533.55+335234.4&Star&	Opt.+NIR&10.88&0.87								&&No\\        
J111612.37+494211.6&Star&	Opt.+NIR&10.91&0.81								&&No\\
J112632.90+120438.0&Quasar&	 X-ray source&20.39&1.04							&&Yes\\
J114913.1$-$0021136.1 &Star&	Optical 				    &14.71	&0.71	&B   &No \\
J121905.30+210633.5&Extended Source& X-ray source&14.83\tablenotemark{e}&1.16	&&No\\ 
J124127.66+085219.5&Quasar&radio+NIR&16.29&1.49				&&yes\\
J124320.01+010844.0&Star&	 Opt.+NIR&14.30&0.85								&&No\\
J131204.71+064107.6&Quasar&	 Opt.+NIR&16.00&1.94								&&Yes\\              
J134517.99+280116.3&Star&	 Opt.+NIR&14.94&1.45								&&No\\       
J135104.50+064307.1&Star&	 Opt.+NIR&15.94&1.63								&&No\\       
J143008.65+230621.6&Seyfert&	 X-ray source&16.05\tablenotemark{e}&1.47		&&No\\            
J143334.42$-$035626.7&Unknown&	Opt.+NIR&12.03&0.85							&&No\\       
J143505.34+404219.4&M Star\tablenotemark{g}&Opt.+NIR&15.12&1.89				&&No\\        
J145426.43+181951.8&Star&	X-ray source\tablenotemark{h}&12.77 &1.14			&&No\\  
J145434.35+080336.7&Quasar&	 Opt.+NIR&15.83\tablenotemark{e}&2.03				&&Yes\\              
J150017.58+121036.5&Star&	Opt.+NIR&15.99&0.99								&&Yes\\       
J150859.72+265148.2&Star&	Opt.+NIR&11.50&0.84								&&No\\        
J151224.30$-$001830.4&Star&	Opt.+NIR&15.69&1.71								&&No\\     
J153334.69+051311.4&Seyfert&radio+NIR&15.89&1.34				&& Yes\\
J153737.84+472604.0&M Star\tablenotemark{g}& Opt.+NIR&15.43&2.85				&&No\\       
J160731.42+173138.4&Seyfert&radio+NIR&16.53&1.38				&& No\\
J162808.20+634924.5&Quasar&	X-ray source&&1.67									&&      \\     	        
J164016.12+295336.3&Star&	Optical&14.82&0.61								&&Yes\\
J164154.20+151753.5&Extended Source&X-ray source&&1.58							&&\\      
J170511.22+225128.4&Star&	Opt.+NIR&15.77&1.01								&&No\\         
J170707.96+222428.2&M Star\tablenotemark{g}&Opt.+NIR&15.09&2.34				&&No\\        
J172832.91+621622.4&M Star\tablenotemark{g}&radio+NIR&17.10&1.51&& No\\
J175740.20+565508.0&Quasar&	X-ray source&&1.70									&&     \\              
J184147.00+321838.5&Extended Source&X-ray source&&1.41							&&\\   
\enddata
\tablenotetext{a}{Classification is based on FWHM of H$\alpha$ line ($>1000$km sec$^{-1}$ for quasar) and
the i-band absolute magnitude ($M_{i} < -22.0$ for quasar; Schneider et al. 2003).}
\tablenotetext{b}{We use PSF magnitude for point sources and Petrosian magnitude for extended sources.}
\tablenotetext{c}{Also selected as a quasar candidate in SDSS?}
\tablenotetext{d}{These two objects are different from each other. We could not identify which one is the target since they are located very close to each other.}
\tablenotetext{e}{Extended source. We use Petrosian magnitude for extended sources, while we use PSF magnitude for point sources.}
\tablenotetext{f}{Cataclysmic variable}
\tablenotetext{g}{The classification may represent that of a nearby bright star. See text.}
\tablenotetext{h}{A galaxy is located 12.06\arcsec from this object. It is possible that this galaxy is the X-ray source since the position accuracy of $ROSAT$ is worse than 13\arcsec.}
\end{deluxetable}
\clearpage

\begin{deluxetable}{ccccc}
\tabletypesize{\scriptsize}
\setlength{\tabcolsep}{0.05in}
\tablecolumns{5}
\tablewidth{4in}
\tablenum{2}
\tablecaption{Observation summary}
\tablehead{
  \colhead{Date}& 
  \colhead{Weather}&
  \colhead{Seeing(\arcsec)}&
  \colhead{Cand.\#\tablenotemark{a}}&
  \colhead{Wavelength coverage}\\
}
\startdata
2005 Jan 5 & clear 			&3.2 & 18  & $5,000-10,000\AA$\\
2005 Jan 6 & humid 			&2.2 & 2 &\\
2005 Jan 7 & cloudy 			&4   & 6 &\\
\hline
2005 May 9 & partly cloudy 	&3.3    & 11& $4,200-8,400\AA$\\
2005 May 10& cloudy 			&3.3 & 4 &\\
2005 May 12& humid 				&2.8 & 1&\\
2005 May 13& cloudy 			&3.1 & 7 &\\
2005 May 14& clear 				&2.5 & 10&\\
\hline
2006 Jun 17--24&&2.5--3.5& 3\tablenotemark{b}& $4,200-8,400\AA$\\
\hline
2006 Dec 20--25&&2.5--3.5& 2\tablenotemark{b} & $4,200-8,400\AA$\\
\enddata                                         
\tablenotetext{a}{Number of observed candidates}
\tablenotetext{b}{Excluded low-{\it b} objects that were the main targets of those observation runs.}
\end{deluxetable}
\clearpage

\begin{deluxetable}{ccccccccccc}
\tablecolumns{11}
\tabletypesize{\scriptsize}
\rotate
\setlength{\tabcolsep}{0.03in}
\tablewidth{8in}
\tablenum{3}
\tablecaption{Properties of quasars}
\tablehead{
  \colhead{SNUQSO Name} & 
  \colhead{$z$} &
  \colhead{$FWHM_{H\alpha}$} &
  \colhead{$FWHM_{H\beta}$} &
  \colhead{$Flux_{H\alpha}$} &
  \colhead{$Flux_{H\beta}$}&
  \colhead{$M_{BH}$}&
  \colhead{SDSS?\tablenotemark{a}}&
  \colhead{$R{^*}\tablenotemark{b}$}&
  \colhead{$M_g$}&
  \colhead{$M_i$}
  \\
  &&${\rm [km~s^{-1}]}$&${\rm [km~s^{-1}]}$&$[\times 10^{-15}{\rm erg~s^{-1}cm^{-2}}]$&$[\times 10^{-15}{\rm erg~s^{-1}cm^{-2}]}$&$[\times 10^6 M_{\sun}]$&&&&\\
}
\startdata
J003236.59$-$091026.2 &$0.092\pm0.001$&	$1856\pm307$	&$2022\pm568$	&	$493\pm62$	&$152\pm32$		&	$25.7^{+9.3}_{-8.7}$	&N						&$4.8$				&-22.75		&-23.16	       \\
J102700.00+390804.2 &$0.144\pm0.001$& 	$2645\pm356$	&$1651\pm678$	&	$104\pm13$ &$12.5\pm4.6$		&	$33.1^{+11.6}_{-10.7}$	&Y						&$\lesssim 3.1$	&-21.80		&-22.37	\\
J103858.44+401058.4 &$0.091\pm0.002$& 	$1267\pm1188$	&				&	$7.49\pm5.89$ &									&	$1.04^{+2.08}_{-2.07}$	&Y		&$\lesssim 2.5$	&-20.99		&-22.01	\\
J104420.11+024819.8 &$0.101\pm0.001$&	$1289\pm536$	&$987\pm480$	&	$80.2\pm23.9$	&$8.23\pm3.08$		&	$4.99^{+4.47}_{-4.42}$		&N				&$10.3$			&-21.23		&-22.10	      \\
J104505.17+084339.1 &$0.123\pm0.001$&  $3776\pm627$	&$4854\pm2122$	&	$359\pm58$ &$95.2\pm41.0$		&	$116^{+47}_{-45}$	&Y							&	$13.0$			&-22.00		&-22.83	      \\
J112632.90+120438.0 &$0.977$        & 					&				&								&									&			&Y	&$\lesssim 1.1$		&-27.50		&-27.65	\\
J124127.66+085219.5 &$0.129\pm0.002$&	$1470\pm470$	&				&	$31.1\pm8.9$	&									&$5.16^{+3.64}_{-3.58}$	&N	&$2.2$				&-22.00		&-22.65	       \\
J131204.71+064107.6 &$0.239\pm0.001$&  $7402\pm608$	&$8825\pm1924$	&	$677\pm55$ &$124\pm27$		&	$1320^{+360}_{-320}$	&N							& $\lesssim 1.5$	&-23.73		&-24.02	\\
J143008.65+230621.6 &$0.079\pm0.001$&  $4674\pm1159$	&$8506\pm2383$	&	$110\pm27$ &$36.1\pm10.0$		&	$58.1^{+32.3}_{-32.0}$	&Y						&$\lesssim 3.5$		&-20.29		&-21.57	\\
J145434.35+080336.7 &$0.131\pm0.001$&	$2391\pm540$	&$2574\pm613$	&	$393\pm84$ &$46.7\pm10.6$		&	$50.6^{+26.3}_{-25.5}$	&N					&$10.9$				&-21.98		&-22.78	      \\
J153334.69+051311.4 &$0.038\pm0.001$&	$4446\pm531$	&$5757\pm1699$	&	$169\pm25$	&$20.6\pm6.0$		&$31.6^{+10.4}_{-9.5}$		&N					&$2.5$				&-19.43		&-20.24	       \\
J160731.42+173138.4 &$0.085\pm0.001$&	$1459\pm315$	&$716\pm670$	&	$59.5\pm7.3$	&$7.02\pm6.83$		&$4.48^{+2.20}_{-2.12}$		&N				&$5.8$				&-20.01		&-21.44	       \\
J162808.20+634924.5 &$0.104\pm0.001$& 	$2103\pm420$	&$1703\pm562$	&	$82.2\pm15.1$ &$18.0\pm5.4$		&	$12.9^{+6.0}_{-5.8}$	&N					&$\lesssim 2.0$	&-21.52\tablenotemark{c}&\\
J175740.20+565508.0 &$0.174\pm0.001$& 	$2269\pm869$	&$2769\pm2031$	&	$44.6\pm16.1$ &$17.3\pm12.2$		&	$18.5^{+15.5}_{-15.3}$	&N				&$\lesssim 4.9$	&-21.73\tablenotemark{c}&\\
\enddata
\tablenotetext{a}{Also discovered as a quasar by SDSS?, Y=Yes, N=No}
\tablenotetext{b}{Radio loudness}
\tablenotetext{c}{$B-$band magnitude (Vega)}
\end{deluxetable}
\clearpage

\begin{deluxetable}{lccc}
\tablecolumns{4}
\tabletypesize{\scriptsize}
\tablewidth{3.5in}
\tablenum{4}
\tablecaption{Efficiency for each method}
\tablehead{
  \colhead{Method}&
  \colhead{Observed}&
  \colhead{Discovered}&
  \colhead{Efficiency}
\\
}
\startdata
Optical Multiple Color		&17	&1	&$5.9\%$\\
Optical+NIR				&17	&3	&$17.6\%$\\
Radio+NIR					&5	&4	&$80.0\%$\\
X-ray							&10	&6	&$60.0\%$\\
Qorg								&6	&0	&$0.0\%$\\
\enddata
\end{deluxetable}
\clearpage

\begin{deluxetable}{cccccc}
\tablecolumns{6}
\tabletypesize{\scriptsize}
\tablewidth{4.5in}
\tablenum{5}
\tablecaption{Estimated properties of host galaxies}
\tablehead{ 
  \colhead{SNUQSO Name}&
  \colhead{$\sigma$}& 
  \colhead{$M_g$}&                      
  \colhead{$m_i$}&                      
  \colhead{$\theta_{50}$\tablenotemark{a}}&
  \colhead{Petro $i-$Fiber $i$\tablenotemark{b}}
  \\    
  &$[{\rm km~s^{-1}}]$&[mag]&[mag]&[\arcsec]&(measured),[mag]\\
}
\startdata
J003236.59$-$091026.2 	&132	&	-19.1	&	17.7	&	1.3		&	\\    
J102700.00+390804.2 	&141& 	-19.4& 	18.2			& 	1.2		&17.7	\\                                                                
J103858.44+401058.4 	&60 &		-16.2& 	20.4		& 	0.3		&16.3	\\
J104420.11+024819.8 	&88 &	-17.6&	19.4			&	0.6		&	\\     
J104505.17+084339.1		&192 &	-20.5& 	16.7			& 	2.7		&\\
J112632.90+120438.0		&       &&						&		&	\\   	                   
J124127.66+085219.5		&89	&	-17.7&	19.9			&	0.5		&17.1	\\      
J131204.71+064107.6		&353 &	-22.7& 	16.0			& 	6.0		&	\\
J143008.65+230621.6		&162	& 	-21.5& 	16.4		& 	2.8		&16.5	\\
J145434.35+080336.7		&157 &	-19.7& 	17.7			& 	1.7		&17.2	\\  
J153334.69+051311.4		&139&	-19.3&	15.5			&	3.9		&16.9	\\      
J160731.42+173138.4		&86&	-17.5&	19.1			&	0.6		&17.3	\\       
J162808.20+634924.5	&111& 	-20.1& 	18.4				& 	1.0	&		\\
J175740.20+565508.0	&122& 	-18.8& 	19.2				& 	0.8	&	\\    
\enddata       
\tablenotetext{a}{Half light radius}                                            
\tablenotetext{b}{We list only objects with visible hosts.}                          
\end{deluxetable}
\clearpage

\end{document}